\documentclass[twocolumn]{aastex63}
\usepackage{times}
\usepackage{amsmath}
\usepackage{textcomp}
\usepackage{amssymb}
\usepackage{natbib}
\usepackage{float}
\usepackage{color}
\usepackage{graphicx}
\usepackage{epstopdf}
\newcommand{\Msun}{\ensuremath{M_{\odot}}}
\newcommand{\lum}{erg\,s$^{-1}$}

\newcommand{\phflux}{\mbox{${\rm \, ph \,\, cm^{-2} \, s^{-1}}$}}

\newcommand{\gm}{$\gamma$}
\newcommand{\OIII}{[O{\sevenrm\,III}]}

 \font\sevenrm=cmr7 scaled 1000

\submitjournal{ApJL}

\shorttitle{Gamma-ray Emission of FR0 Galaxies}
\shortauthors{Vaidehi S. Paliya}

\begin{document}
\title{A New Gamma-ray Emitting Population of FR0 Radio Galaxies}

\correspondingauthor{Vaidehi S. Paliya}

\author[0000-0001-7774-5308]{Vaidehi S. Paliya}
\affiliation{Aryabhatta Research Institute of Observational Sciences (ARIES), Manora Peak, Nainital 263001, India}

\email{vaidehi.s.paliya@gmail.com}

\begin{abstract}
The enigmatic class of Fanaroff$-$Riley type 0 (FR0) radio galaxies is emerging as the missing link between the faint yet numerous population of compact radio sources in nearby galaxies and the canonical Fanaroff$-$Riley classification scheme. This letter reports the first \gm-ray identification of three FR0 galaxies above 1~GeV using more than a decade of the Fermi Large Area Telescope observations.  A cumulative \gm-ray emission at $>$5$\sigma$ significance was also detected from the \gm-ray unresolved FR0 sources using the stacking technique, suggesting the FR0 population to be a \gm-ray emitter as a whole. The multi-frequency properties of the \gm-ray detected sources are similar to other FR0s, thus indicating the high-energy radiation to originate from misaligned jets. Given their large abundance, FR0 radio galaxies are proposed as plausible candidates for IceCube-detected neutrinos and the results presented in this letter may provide crucial constraints on their \gm-ray production mechanism and the origin of cosmic neutrinos.
\end{abstract}


\section{Introduction}{\label{sec:Intro}}
Blazars are active galactic nuclei (AGN) hosting closely aligned relativistic jets. They are the most abundant class of extragalactic \gm-ray emitters \citep[][]{2020ApJ...892..105A} due to the flux enhancement caused by Doppler boosting.  This effect is considerably less prominent in their misaligned counterparts, i.e., radio galaxies, though a small number of them are detected in \gm-ray surveys \citep[see, e.g.,][]{2020ApJ...892..105A}. Radio galaxies have been classified as faint edge-darkened Fanaroff$-$Riley type I (FR I) and bright edge-brightened FR II galaxies based on their extended radio morphology \citep[][]{1974MNRAS.167P..31F}.  FR Is are more numerous \gm-ray emitters possibly due to their relatively flat \gm-ray spectra though it is not well understood \citep[cf.][]{2010ApJ...720..912A}.

Recently, a new class of faint yet abundant radio sources has been identified with several observational characteristics similar to FR I and II galaxies. The defining feature of these objects is the lack of extended, i.e., kpc scale, radio emission,  distinguishing them from FR Is and IIs \citep[][]{2009A&A...508..603B,2015A&A...576A..38B}. This emerging source class is termed as `FR0' radio galaxies \citep[][]{2011AIPC.1381..180G}. The nuclear properties of FR 0s, e.g., bolometric AGN power, are found to be similar to that of FR I galaxies and so their hosts, which are typically red early-type galaxies  \citep[cf.][]{2009A&A...508..603B}.

\begin{figure*}[t!]
\hbox{
\includegraphics[scale=0.425]{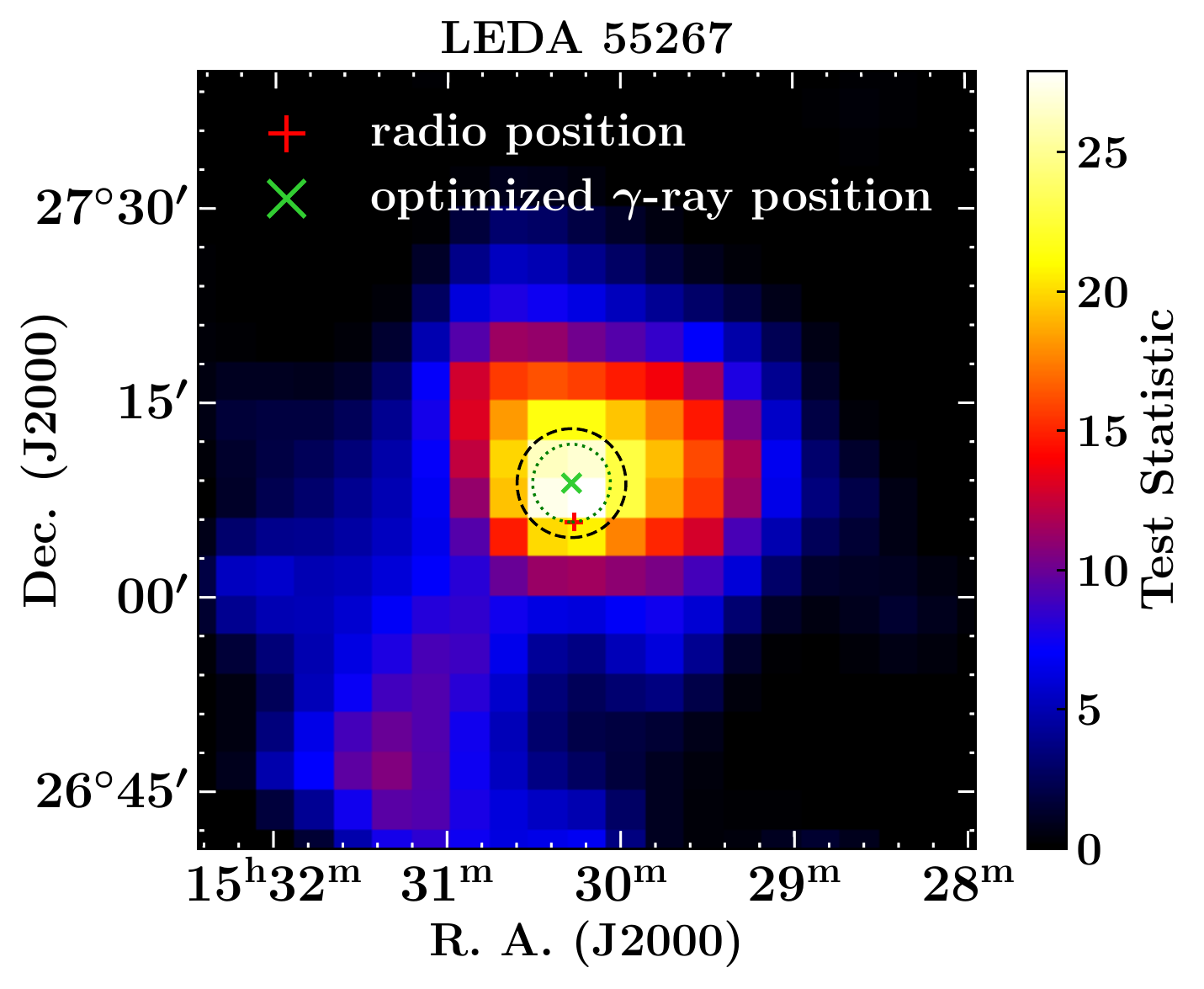}
\includegraphics[scale=0.425]{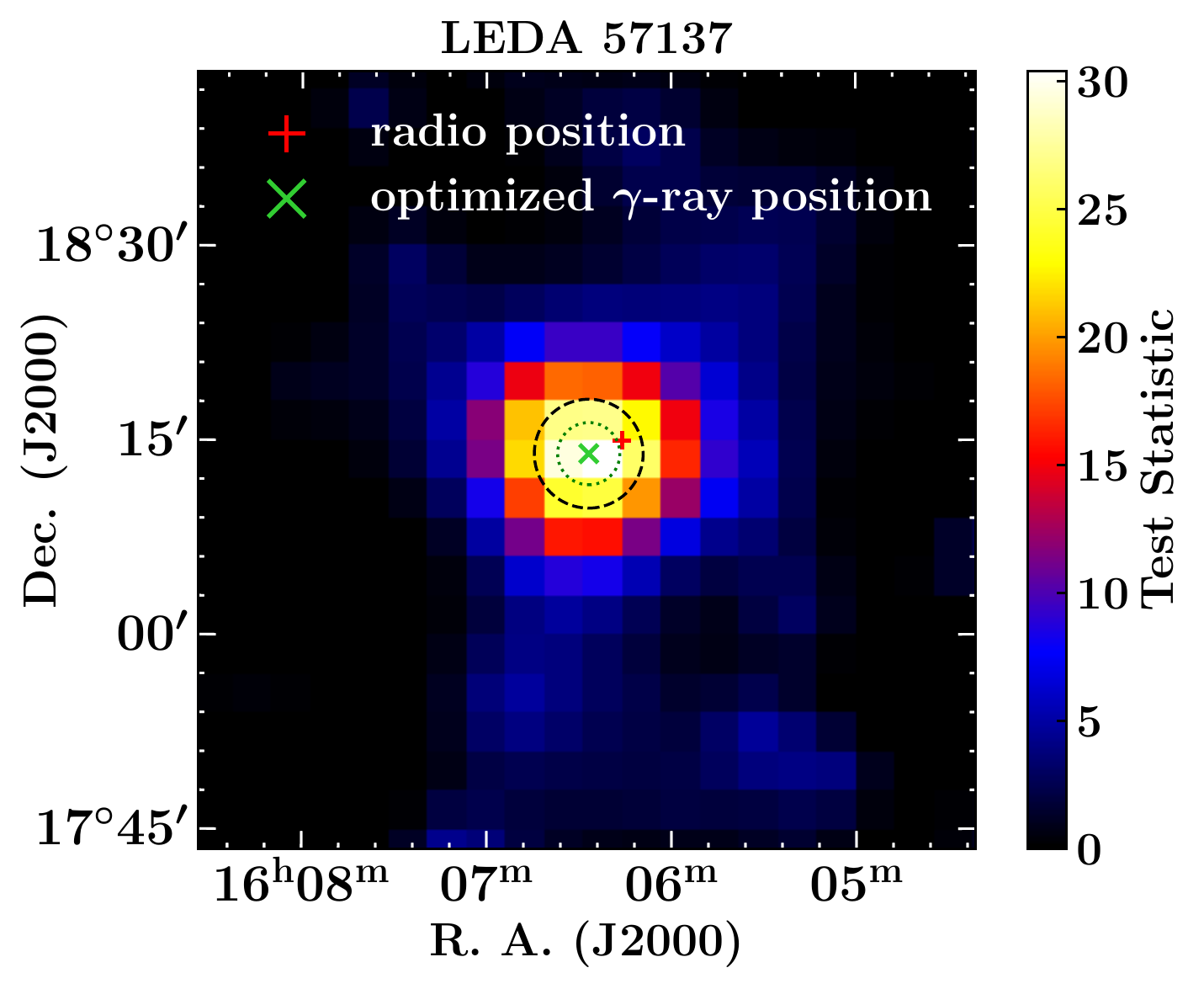}
\includegraphics[scale=0.425]{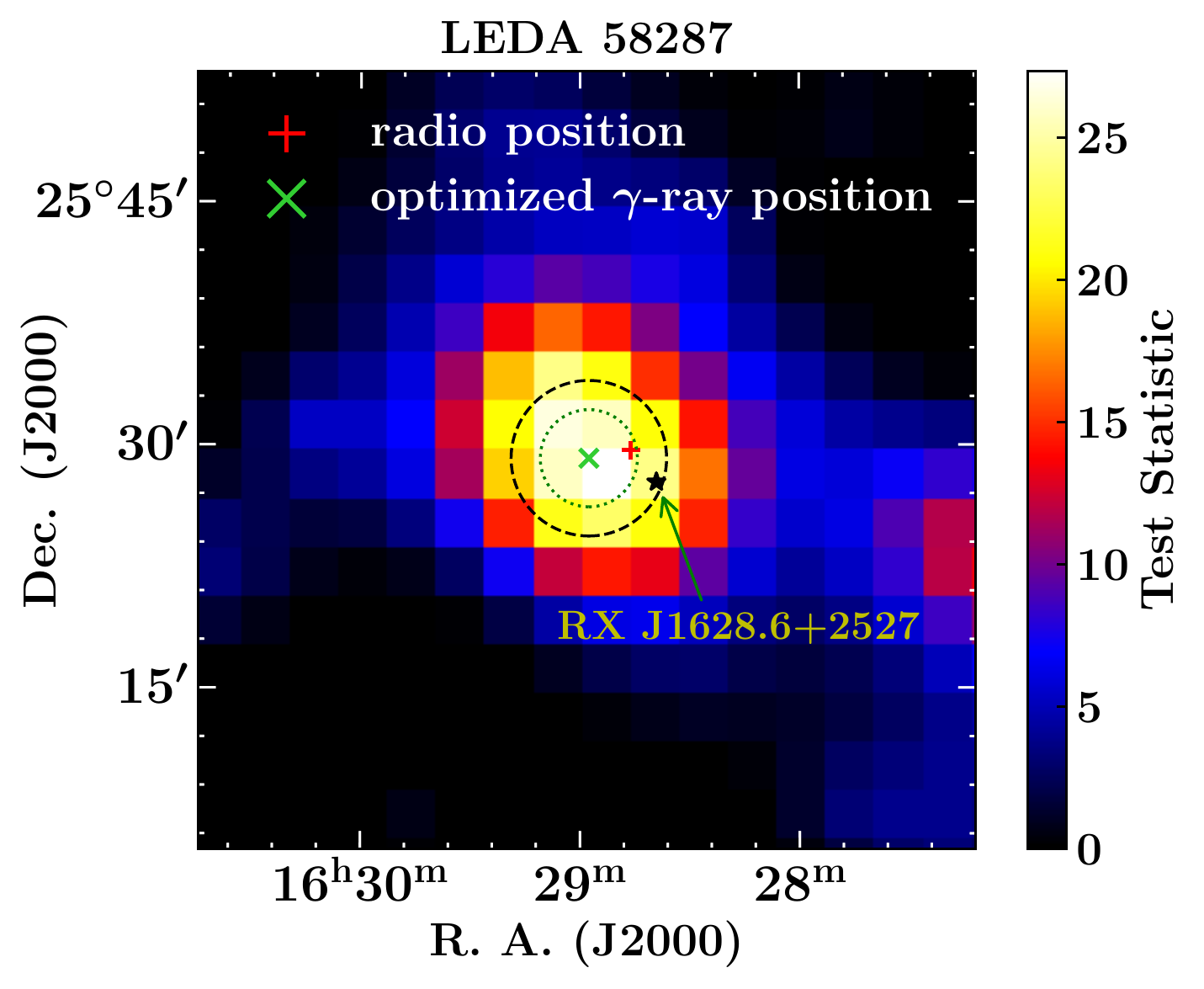}
}
\caption{The residual TS maps of three FR0 radio galaxies detected in the \gm-ray band. Green dotted and black dashed circles represent 68\% and 95\% uncertainty regions, respectively.} \label{fig:ts_map}
\end{figure*}

The possibility that FR0 radio galaxies could be \gm-ray emitters \citep[][]{2016MNRAS.457....2G} has opened up a new dimension in the multi-messenger astronomy, also because FR0 forms the faint yet bulk of radio-loud AGN population. These galaxies have been proposed as promising sources of IceCube detected neutrinos \citep[][]{2018MNRAS.475.5529T} and also as plausible sites of ultra-high-energy cosmic ray acceleration \citep[cf.][]{2021APh...12802564M}. Therefore, a systematic characterization of FR0 galaxies in the \gm-ray band is the need of the hour. Even if individual objects remain below the detection threshold of the Fermi Large Area Telescope (LAT), their cumulative \gm-ray signal could provide meaningful information constraining the proposed hypotheses. It is the primary objective of the presented work.

In this letter, the first \gm-ray detection of three FR0 radio galaxies is reported, including a candidate \gm-ray emitter, along with the identification of this class of radio-loud AGN as a new population of \gm-ray emitters following a stacking analysis technique. The FR0 sample and details of the LAT data reduction are presented in Section~\ref{sec2}. Results are presented in Section~\ref{sec3} and \ref{sec4}. Section~\ref{sec5} is devoted to the discussion and summary of the obtained results. The cosmology parameters used are $H_0=67.8$~km~s$^{-1}$~Mpc$^{-1}$, $\Omega_m = 0.308$, and $\Omega_\Lambda = 0.692$ \citep[][]{2016A&A...594A..13P}. The quoted errors are 1$\sigma$ statistical uncertainties, unless specified.

\section{Data Reduction}\label{sec2}
\citet[][]{2018A&A...609A...1B} presented a catalog of bonafide 108 FR0 objects, called {\it FR0CAT}, using the radio galaxy sample of \citet[][]{2012MNRAS.421.1569B}. The sources included in {\it FR0CAT} have redshift $z\leq0.05$, extended radio emission $\lesssim$5 kpc,  and their optical spectroscopic properties are analogous to low-excitation galaxies. The P8R3 Fermi-LAT data of all 108 FR0s\footnote{The previously reported \gm-ray emitting FR0 galaxy, Tol 1326$-$379 \citep[][]{2016MNRAS.457....2G} is not included in {\it FR0CAT}.} were analyzed to identify potential \gm-ray emitters.

The binned likelihood fitting method was adopted to determine the significance of the \gm-ray signal from the SOURCE class events lying within a circular region of interest (ROI) of radius 10$^{\circ}$ centered at the target AGN. The considered time and energy ranges were 2008 August 4 to 2021 May 5 (MJD 54683$-$59339) and 1$-$1000 GeV, respectively. The minimum energy was chosen as 1 GeV because the Fermi-LAT point spread function significantly improves above this energy \citep[cf.][]{2013arXiv1303.3514A} allowing a better source localization and suppressing the diffuse background emission brighter at MeV energies \citep[e.g.,][]{2016ApJS..223...26A}. All \gm-ray sources present in the second data release of the fourth catalog of the Fermi-LAT detected sources \citep[4FGL-DR2;][]{2020ApJS..247...33A,2020arXiv200511208B} and lying within 15$^{\circ}$ from the target, along with the recommended diffuse background models\footnote{https://fermi.gsfc.nasa.gov/ssc/data/access/lat/BackgroundModels.html}, were used to model the \gm-ray sky. Spectral parameters of sources lying within the ROI were kept free during the fit to derive the maximum likelihood test statistic \citep[TS;][]{1996ApJ...461..396M}. It is defined as TS =  $2\log(\mathcal{L}$), where $\mathcal{L}$ denotes the ratio of the likelihood values with and without a point source at the position of interest. Since the covered period was larger than that adopted in the 4FGL-DR2 catalog, TS maps were generated to identify unmodeled \gm-ray excesses present in the data but not in the model. Once all such excesses were taken into account, a final likelihood fitting was carried out to compute the detection significance and spectral parameters of FR0 radio galaxies. Sources with TS$>$25 ($\sim$4.2$\sigma$) were considered as \gm-ray detected ones.

\begin{deluxetable*}{lcccccccccc} 
\tablewidth{0pt} 
\tablecaption{Parameters associated with the FR0 radio galaxies detected in the energy range 1$-$1000 GeV.\label{tab:basic}}
\tablewidth{0pt}
\tablehead{
\colhead{Name} & \colhead{$z$} & \multicolumn{2}{c}{Optimized \gm-ray position} & \colhead{$R_{95\%}$} & \colhead{TS} & \colhead{$F_{\gamma}$} & \colhead{$\Gamma_{\gamma}$} & \colhead{$L_{\gamma}$} & \colhead{$L_{\rm \OIII}$} & \colhead{$M_{\rm BH}$}\\
\colhead{} & \colhead{} & \colhead{hh mm ss.ss} & \colhead{dd mm ss.s} & \colhead{(deg.)} & \colhead{} & \colhead{(10$^{-10}$ \phflux)} & \colhead{} & \colhead{(10$^{42}$ \lum)} & \colhead{(\lum)} & \colhead{(\Msun)}
}
\startdata
LEDA 55267	  & 0.033 & 15 30 16.99  & +27 08 52.1   & 0.07 & 29 & 1.07$\pm$0.31 & 2.23$\pm$0.23 & 2.0$\pm$0.8  & 5.1$\times$10$^{39}$ & 1.6$\times$10$^{8}$\\
LEDA 57137       & 0.037 & 16 06 26.81  & +18 13 58.1    & 0.07 & 30 & 1.25$\pm$0.32 & 2.22$\pm$0.12 & 3.0$\pm$1.0 & 4.8$\times$10$^{39}$& 5.0$\times$10$^{8}$\\
LEDA 58287      & 0.04   & 16 28 57.58  & +25 29 10.7   & 0.08 & 27 & 0.99$\pm$0.29 & 2.04$\pm$0.20 & 3.9$\pm$1.7  & 4.5$\times$10$^{39}$& 3.2$\times$10$^{8}$\\
\enddata
\tablecomments{$R_{95\%}$ is the 95\% uncertainty radius in the optimized \gm-ray position. Last three columns show the \gm-ray and \OIII~line luminosities, and black hole mass, respectively.}
\end{deluxetable*}

A likelihood profile stacking technique was employed to derive the overall detection significance of the \gm-ray undetected sources. This method has already been used for various studies, e.g., upper limits on dark matter interaction, extreme blazars, and star-forming galaxies, \citep[cf.][]{2011PhRvL.107x1302A,2019ApJ...882L...3P,2020ApJ...894...88A}. The basic assumption of this approach is that the considered source population can be characterized by an average photon flux ($F_{\gamma}$) and photon index ($\Gamma_{\gamma}$) when their \gm-ray spectra are modeled with a power-law. In particular, a grid of $F_{\gamma}$ and $\Gamma_{\gamma}$ was created, and TS values were calculated at each grid point, effectively generating a TS profile for each FR0 galaxy. Since log-likelihood is additive, TS profiles of all sources were stacked and scanned to determine the cumulative TS peak and associated spectral parameters. Further details of this technique can be found in \citet[][]{2019ApJ...882L...3P}.

 \begin{figure*}[t!]
\hbox{
\includegraphics[scale=0.3]{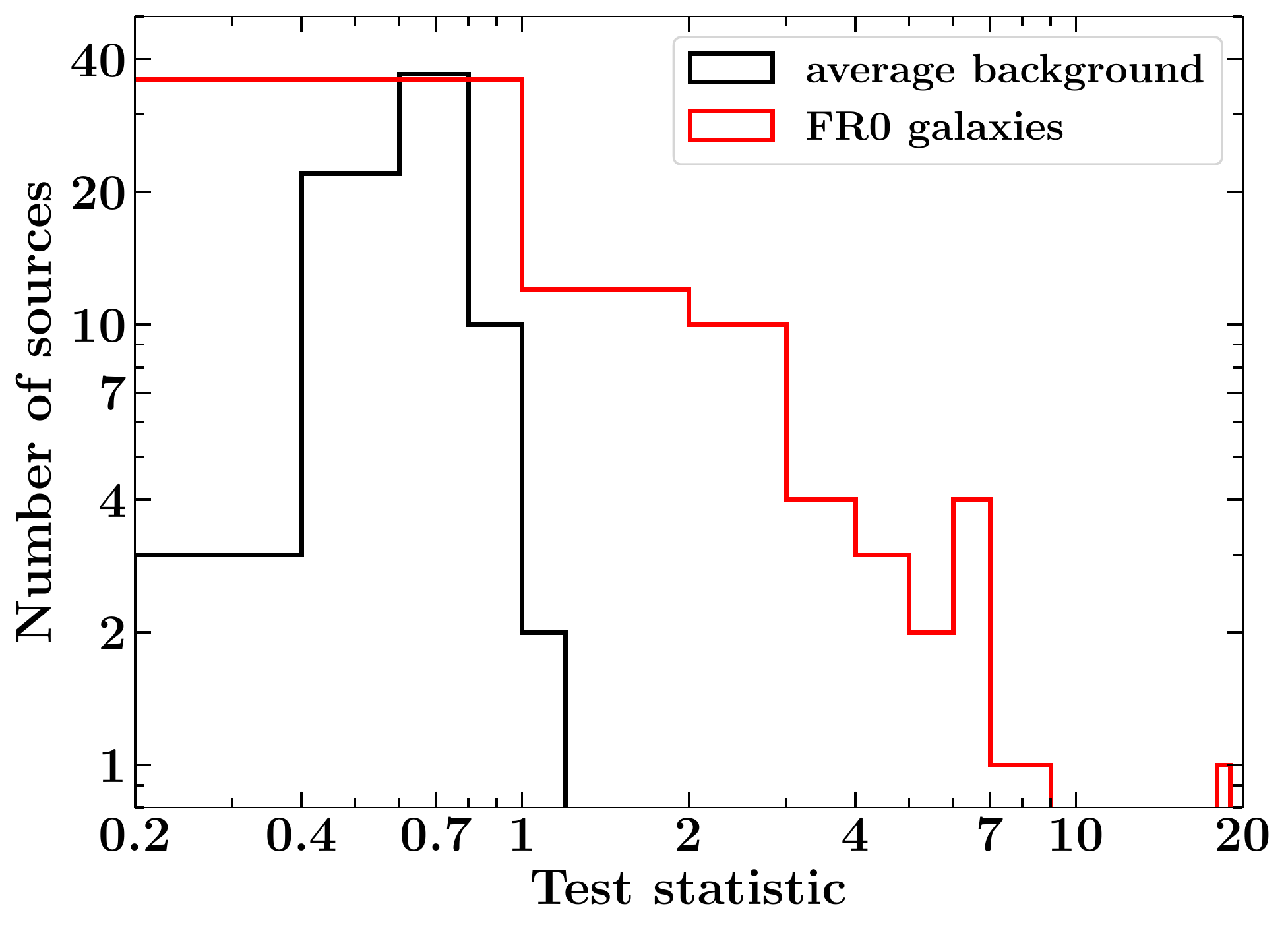}
\includegraphics[scale=0.275]{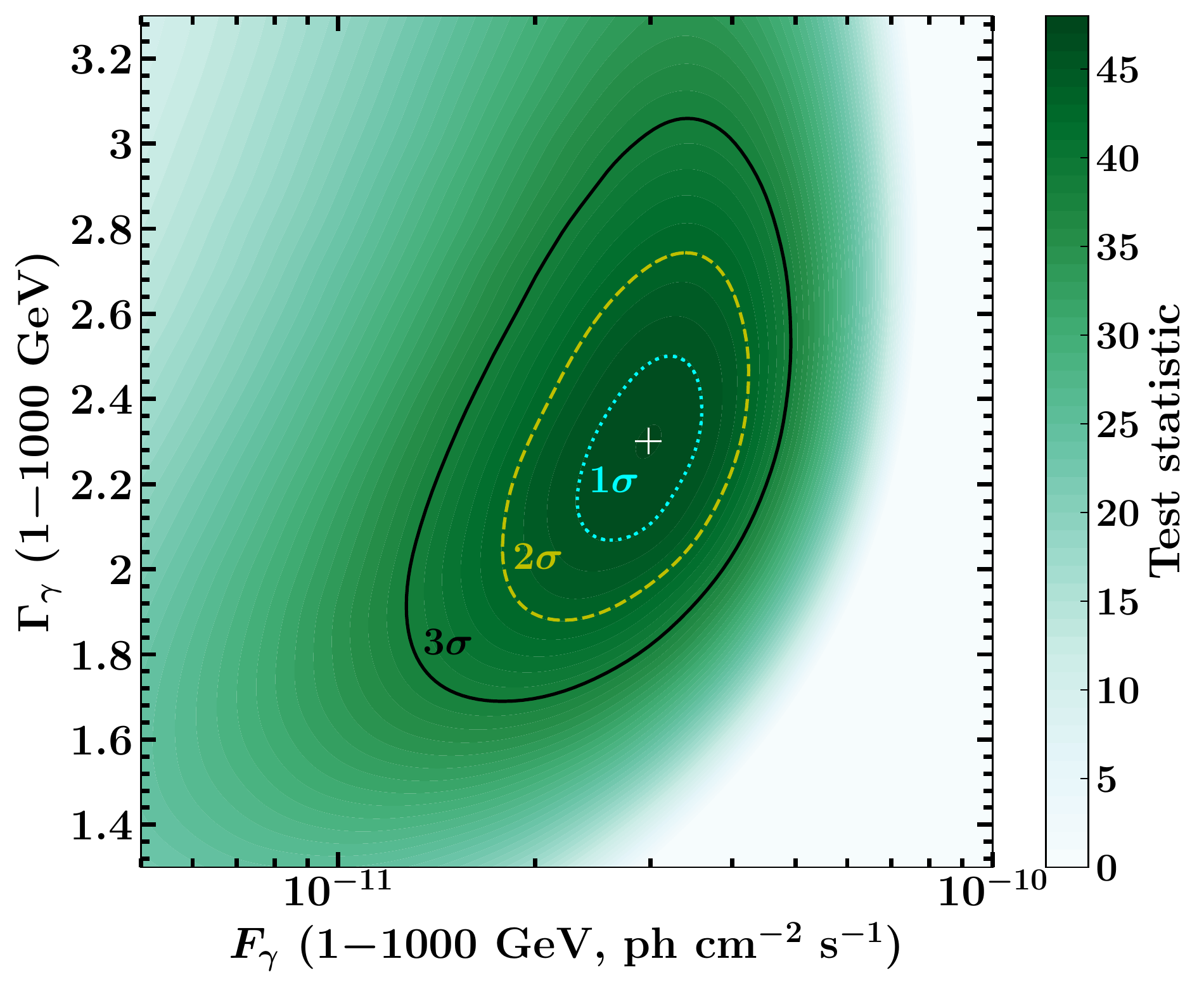}
\includegraphics[scale=0.365]{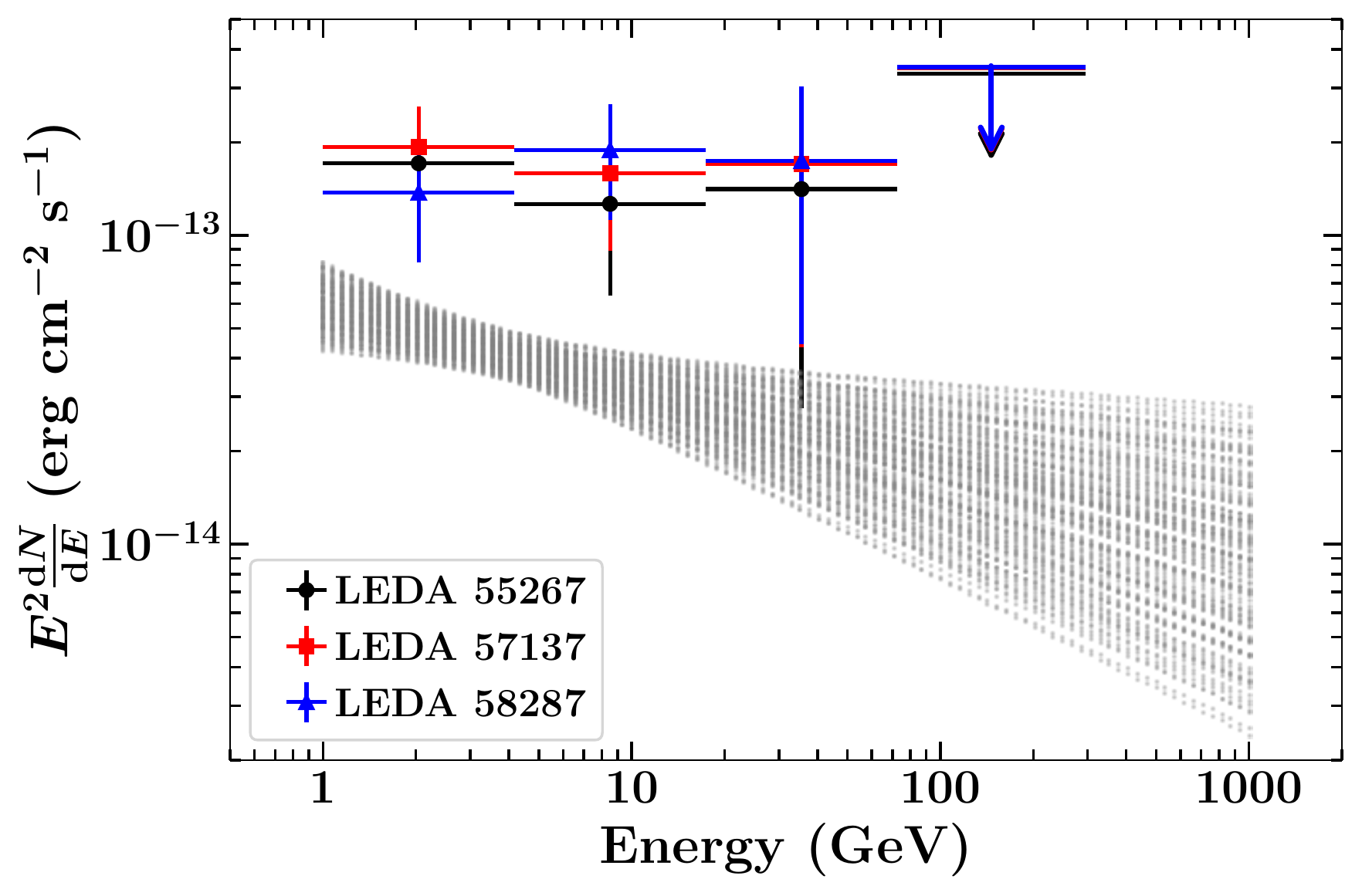}
}
\caption{Left: Histogram of the TS values for 74 FR0 galaxies (red) selected for the stacking analysis. The black histogram represents the average TS of the 100 iterations carried out to randomly pick 74 TS$>$0 background regions from the original pool of 500 empty sky positions. Middle: The background subtracted combined TS profile of 74 FR0 galaxies. Confidence contours are at 1$\sigma$, 2$\sigma$, and 3$\sigma$ levels. Right: Stacked bow-tie spectrum of the \gm-ray undetected FR0 galaxy sample. Spectra of three \gm-ray detected objects are also shown.  Flux upper limits were computed at 95\% confidence level.} \label{fig:stack}
\end{figure*}

\section{New Gamma-ray Emitting FR0 Galaxies}\label{sec3}
The LAT data analysis of 108 FR0 radio galaxies has revealed a significant \gm-ray emission from 3 objects above 1 GeV.  The residual TS maps of these sources are shown in Figure~\ref{fig:ts_map} along with their radio and optimized \gm-ray positions. The derived spectral parameters are provided in Table~\ref{tab:basic}. To ascertain the association of the \gm-ray source with the FR0s,  1.4 GHz flux density maps from the Faint Images of the Radio Sky at Twenty-Centimeters (FIRST) survey were inspected. Two out of three FR0 galaxies, LEDA 55267 and LEDA 57137, were found to be the brightest radio sources within the 95\% uncertainty region associated with the optimized \gm-ray position. The case of LEDA 58287 is complex due to the presence of a BL Lac object, RX J1628.6+2527 ($z=0.22$), at the edge of the 95\% uncertainty region (Figure~\ref{fig:ts_map}, right panel). The BL Lac is slightly brighter at 1.4 GHz with flux density $\sim$46 mJy compared to LEDA 58287 ($F_{\rm 1.4~GHz}\sim$25 mJy). However, the FR0 galaxy is closer to the optimized \gm-ray position being located inside 68\% uncertainty region while the BL Lac object is outside of it. With the collection of more LAT data, it will be possible to further reduce the uncertainty region size and identifying the true counterpart of the \gm-ray source. Therefore, the FR0 radio galaxy LEDA 58287 can be considered as a candidate \gm-ray emitter.

 \begin{figure*}[t!]
\hbox{\hspace{2.4cm}
\includegraphics[scale=0.38]{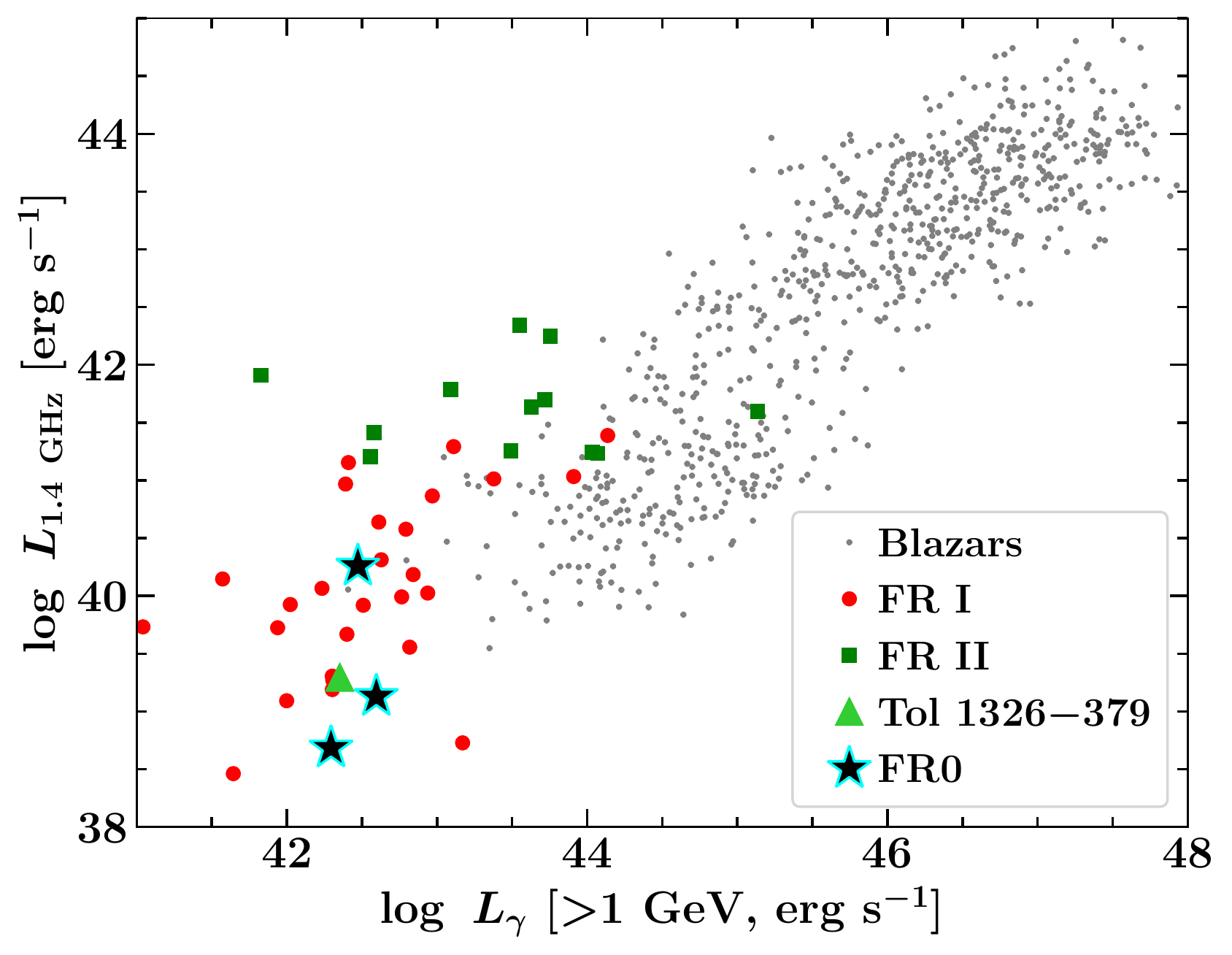}
\includegraphics[scale=0.38]{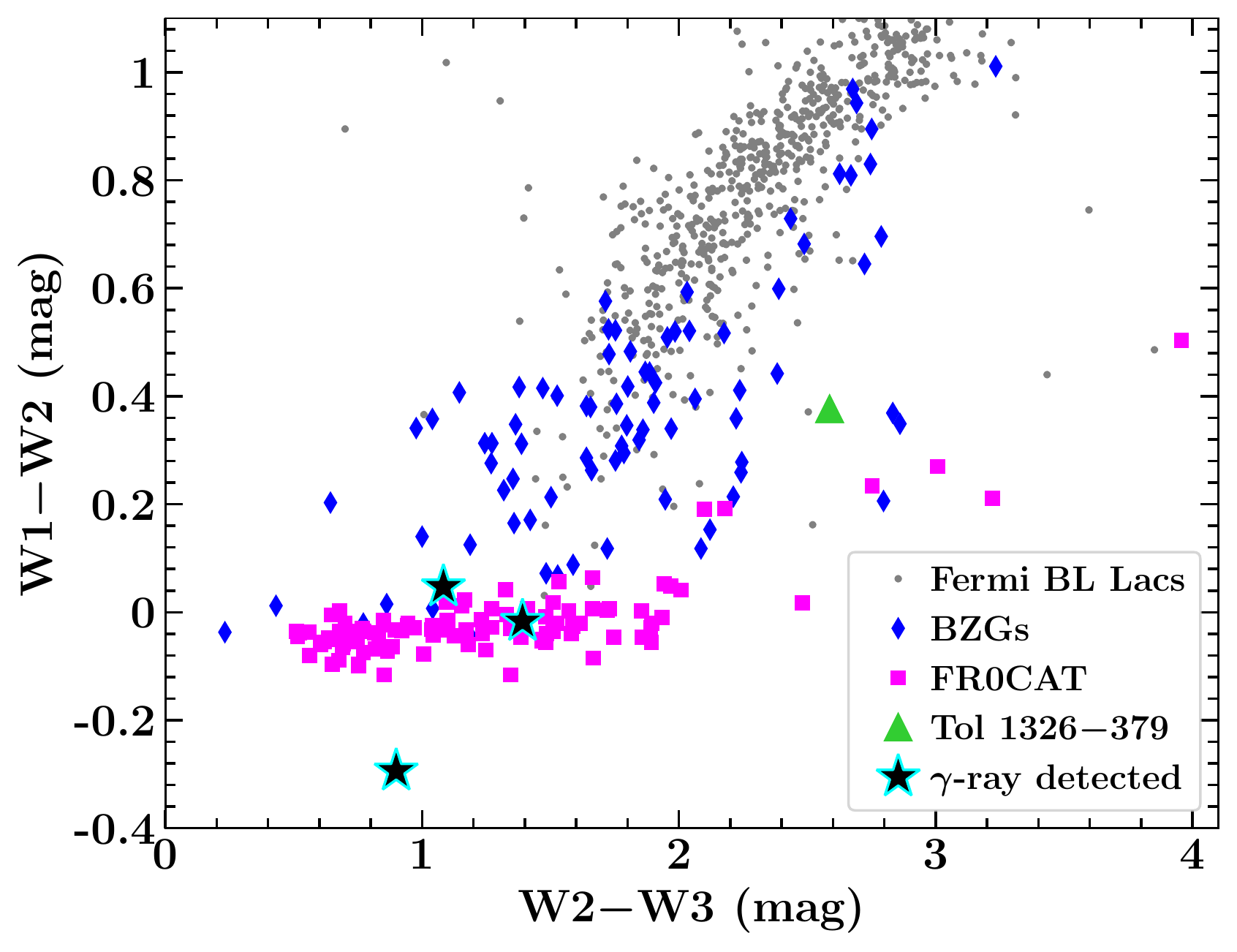}
}
\caption{Left: Radio luminosity as a function of the \gm-ray luminosity plotted for blazars and radio galaxies. Right: WISE color-color diagram showing the regions populated by blazars, BZGs, and FR0 radio galaxies. The \gm-ray detected FR0s are highlighted with black stars.  In both plots, the previously reported \gm-ray emitting FR0 galaxy, Tol 1326$-$379, is also shown.} \label{fig:multi}
\end{figure*}

\section{Stacking of the Undetected Population}\label{sec4}
There are 77 FR0 sources present in the sample that have TS$>$0, thus indicating the FR0 population to be a faint \gm-ray emitter as a whole. To quantify the significance of the \gm-ray signal, the stacking pipeline was employed on 74 objects after excluding the \gm-ray detected ones. Furthermore, to take into account the background radiation, 500 empty sky positions at high galactic latitude ($|b|>10^{\circ}$) were randomly selected that do not lie within 95\% uncertainty regions of any 4FGL-DR2 objects and the whole process was repeated.  Since the background contamination can be non-uniform, 74 empty sky positions with TS$>$0 were randomly picked from the pool\footnote{Since 74 FR0 galaxies picked for stacking were chosen based on TS$>$0 condition, the same was applied to select 74 empty sky positions.} and their stacked TS profile was subtracted from that obtained for the FR0 galaxy sample. This allowed the measurement of the TS peak from the background-subtracted TS profile of FR0 galaxies. This step was repeated 100 times by randomly picking 74 TS$>$0 empty sky positions in each step. The mean of the distribution is 48 and the TS peak was found to be $>$25 in 98 out of 100 iterations. The average histogram of the 74 TS$>$0 background regions picked in 100 iterations is shown in the left panel of Figure~\ref{fig:stack}. The TS histogram of 74 FR0 galaxies is overplotted in the same panel and a comparison of the two highlights the significance of \gm-ray emission in the selected sample of FR0s above the background. The background-subtracted TS profile of one of the iterations is shown in Figure~\ref{fig:stack} (middle panel). In the energy range of 1$-$1000 GeV, the average photon flux and photon index for the FR0 source population are $\langle F_{\gamma}\rangle=2.99^{+0.56}_{-0.54}\times 10^{-11}$ \phflux~and $\langle\Gamma_{\gamma}\rangle=2.30^{+0.18}_{-0.20}$, respectively. In the right panel of Figure~\ref{fig:stack}, the stacked \gm-ray spectrum, as well as spectra of the detected FR0 galaxies, are shown.

\section{Discussion and Summary}\label{sec5}
With the \gm-ray detection of three FR0 galaxies, including a candidate source, and a significant cumulative \gm-ray signal from a major fraction of {\it FR0CAT} objects revealed by the stacking analysis, the FR0 source population can now be considered as a \gm-ray emitter, similar to, e.g., blazars and FR I and II galaxies.

Using the \OIII~line luminosity ($L_{\rm \OIII}$) reported in {\it FR0CAT}, the bolometric luminosity of the \gm-ray detected sources can be computed following $L_{\rm bol}=3500L_{\rm \OIII}$ \citep[Table~\ref{tab:basic};][]{2004ApJ...613..109H}. The accretion rate in Eddington units is then derived as $\dot L =L_{\rm bol}/L_{\rm Edd}$ using the black hole mass listed in {\it FROCAT} and it turned out to be $<$ 10$^{-3}$ for all of them, a value typical of low-excitation galaxies and similar to that found for other FR0 sources \citep{2012MNRAS.421.1569B,2018A&A...609A...1B}. Furthermore, a comparison of the $k$-corrected radio and \gm-ray luminosities of these objects with 4FGL-DR2 blazars and FR I and II radio galaxies \citep[][]{2020APh...11602393A} is shown in the left panel of Figure~\ref{fig:multi}. All three FR0s lie in the low-luminosity tail of the blazar distribution. \citep[see also][]{2016MNRAS.457....2G}.  Interestingly, this gives the impression that these objects may belong to the low-luminosity BL Lac population, i.e., their jets might be viewed at small angles. The similar large-scale environments of FR0s and BL Lac sources also indicate a possible connection between these two populations \citep[][]{2020ApJ...900L..34M}. This was tested by inspecting the location of the \gm-ray emitting FR0 galaxies in the Wide-field Infrared Survey Explorer (WISE) color-color diagram where blazars are known to occupy a distinct region \citep[][]{2011ApJ...740L..48M}. As can be seen in Figure~\ref{fig:multi} (right panel), FR0 galaxies, including the \gm-ray emitting ones, populate a region substantially far from that occupied by blazars.  This observation suggests that the infrared emission in FR0 galaxies is probably not dominated by the jet. Moreover,  this finding also supports the fact that these FR0 sources may not host a beamed relativistic jet.  This is further reinforced by plotting the ``BL Lac-galaxy dominated" (BZGs\footnote{BZGs are proposed as BL Lac objects with a dominant host galaxy emission overwhelming the nuclear radiation \citep[][]{2015Ap&SS.357...75M}.}) sources \citep[][]{2015Ap&SS.357...75M}. Several of these can be sources with weak or misaligned jets, and in fact they appear to occupy a region in the WISE color-color plot that is intermediate between BL Lacs and FR0s (Figure~\ref{fig:multi}, right panel).

Another observational evidence corroborating the misaligned nature of the \gm-ray detected FR0 galaxies comes from the ratio of the \OIII~and 2$-$10 keV luminosities, $R_{\rm \OIII/X} = \log(L_{\rm \OIII}/L_{\rm 2-10~keV})$. Since the X-ray emission is beamed in BL Lac objects, this ratio is found to be small \citep[$\sim-3.3$, cf.][]{2015A&A...580A..73C} compared to FR0 radio galaxies \citep[$\sim-1.7$, e.g.,][]{2009MNRAS.396.1929H,2018MNRAS.476.5535T}. Using the Swift X-ray telescope observations, $R_{\rm \OIII/X}$ was determined as $-$1.7,  $-$1.9, and $-$1.4 for LEDA 55267, LEDA 57137, and LEDA 58287,  respectively, similar to other FR0 galaxies. Altogether, the \gm-ray emission in FR0 systems is likely to be originated from misaligned jets.

The \gm-ray production mechanism in radio galaxies is still a matter of debate. Viewed at large jet angles, their \gm-ray emission is expected to be suppressed due to relatively smaller Doppler boosting. However, many such objects have been discovered with Fermi-LAT and \gm-ray flaring activities have also been reported \citep[see, e.g.,][]{2018ApJ...860...74T}. A one-zone synchrotron self Compton model is commonly adopted to explain the \gm-ray emission, however, it fails to explain the TeV radiation \citep[cf.][]{2008MNRAS.385L..98T}. More complex radiative scenarios have been proposed such as spine-shear topology \citep[e.g.,][]{2008MNRAS.385L..98T}, multi-zone emission \citep[cf.][]{2003ApJ...594L..27G,2008A&A...478..111L}, and/or various magnetic reconnection driven events \citep[][]{2010MNRAS.402.1649G}. 

Hadronic models based on $p$\gm~interaction and proton synchrotron process have also been invoked to explain the \gm-ray emission \citep[e.g.,][]{2004A&A...419...89R}. Since many of the observational properties of FR0 galaxies are similar to FR Is, their \gm-ray production mechanism might also be similar. In particular, hadronic processes also predict relativistic jets of these sources to be sites of ultra-high-energy cosmic-ray acceleration and neutrinos \citep[cf.][]{2017JCAP...12..017B}. Indeed, the lack of point source detection in IceCube events hints at a numerous population of faint objects as a source of the diffuse neutrino flux. Interestingly, FR0 radio galaxies belong to one such type of astrophysical population leading \citet[][]{2018MNRAS.475.5529T} to propose them as a promising source of IceCube detected neutrinos. In this regard,  the \gm-ray spectral parameters derived for both individually and cumulatively \gm-ray detected FR0 radio galaxies (Table~\ref{tab:basic}) would be crucial to constrain the proposed \gm-ray emission mechanisms and origin of cosmic neutrinos.

\acknowledgements
The author thanks the journal referee for constructive criticism.

\bibliographystyle{aasjournal}

\end{document}